\documentclass[preprint,aps,tightenlines,floatfix,showpacs]{revtex4}
\usepackage{graphics}
\usepackage{bm}
\usepackage{amsmath}
\usepackage{amssymb}
\begin{document}

\title{Components of an
algebraic solution of the multichannel problem 
of low-energy n-${}^{12}$C scattering plus sub-threshold 
(${}^{13}$C) states}
%%%%%  Authors, affiliations  %%%%%%
\author{K. Amos}
\email{amos@physics.unimelb.edu.au}
\affiliation{School of Physics, University of Melbourne,
Victoria 3010, Australia}

\author{L. Canton}
\email{luciano.canton@pd.infn.it}
\author{G. Pisent}
\email{gualtiero.pisent@pd.infn.it}
\affiliation{Istituto Nazionale di Fisica Nucleare, sezione di Padova, \\
e Dipartimento di Fisica dell'Universit$\grave {\rm a}$
di Padova, via Marzolo 8, Padova I-35131,
Italia}

\author{J. P. Svenne}
\email{svenne@physics.umanitoba.ca}
\affiliation{
Department of Physics and Astronomy,
University of Manitoba,
and Winnipeg Institute for Theoretical Physics,
Winnipeg, Manitoba, Canada R3T 2N2}

\author{D. van der Knijff}
\email{dirk@unimelb.edu.au}
\affiliation{Advanced Research Computing, Information Division,
University of Melbourne, Victoria 3010, Australia}

\date{today}

%%%%%  Abstract  %%%%%
\begin{abstract}
The effects of components in an assumed model interaction potential,
as well as of the order to which its deformation is taken, upon resonances in
the low-energy cross sections and upon sub-threshold bound states of the 
compound nucleus (${}^{13}$C) are discussed.
\end{abstract}
%%%%%%%%%%%%%%%%%%%%%%%%%%%%%%%%%%%%%%%%%%%%%

\maketitle

%%%%%%%%%%%%%%%%%%%%%%%%%%%%%%%%%%%%%%%%%%%%%%%%%%%%%%%%%%%5

In a recent paper~\cite{Am03}, we specified in detail a multichannel 
algebraic scattering theory (MCAS) for nucleons scattering from a nucleus. 
The approach is noteworthy because it allows a) a systematic determination
of the sub-threshold  bound states and compound resonances and b)
the inclusion in the collective-model excitation of the target of the 
non-local effects due to the Pauli principle.
The theory
was built upon sturmian expansions of a model interaction matrix of
potential functions with the dimensionality of the problem rapidly increasing
with the number of closed and open channels to be taken into consideration.
Thus the first application was of low-energy neutron-${}^{12}$C scattering
since then only three states sensibly were needed in the evaluations; namely
the ground $0^+_1$, $2^+_1$ (4.44), and the $0^+_2$ (7.65) states in ${}^{12}$C.

We chose a collective-model representation for the interaction potential
matrix with deformation taken to second order.  The potential function
involved of a set of operators, and with the parameter values specified in
the Table~\ref{OMparams}, we obtained excellent results for the total elastic  
scattering cross section to 4 MeV excitation. Of particular note was that
both broad and narrow resonances of the correct spin-parity were found
at the correct excitation energies with appropriate widths and maxima
resting upon an smooth and realistic background.
%%%%%%%%%%%%%%%%%%%%%%%%%%%%%%%%%%%%%%%%%%%%%%%%%%%%%%%%%%%%%%%%%%%%%
\begin{table}
\begin{ruledtabular}
\caption
{\label{OMparams}\sf The parameter values defining the base potential}
\begin{tabular}{cccccc}
 & $V_{\rm central}$ & $V_{\ell \ell}$ & $V_{\ell s}$ & $V_{Is}$ \\
\hline
\ \ \ Odd parity &
-49.144 & 4.559 & 7.384 & -4.770 \\
\ \ \ Even\ parity\hspace*{0.4cm}  &
-47.563 & 0.610 & 9.176 & -0.052 \\
\hline
\ \ \ Geometry & \hspace*{0.2cm} $R_0 = 3.09$ fm  \hspace*{0.2cm}  &
\hspace*{0.2cm} $a = 0.65$ fm \hspace*{0.2cm} &
\hspace*{0.2cm}  $\beta_2 = -0.52$\hspace*{0.2cm}  &
\hspace*{2.0cm}\\
\end{tabular}
\end{ruledtabular}
\end{table}
%%%%%%%%%%%%%%%%%%%%%%%%%%%%%%%%%%%%%%%%%%%%%%%%%%%%%%%%%%%%%%%%%%%%%

The same theory can be applied using negative energies. Then 
the resonance finding process discussed in ref.~\cite{Am03}
determined the sub-threshold bound states of the compound nucleus; in our
case ${}^{13}$C.  With the potential parameters listed, agreement
with the sub-threshold bound state spectrum resulted.  Notably we found 
the right number of states, with the right spin-parities and at energies 
in quite good agreement with observation. But it was crucial to allow
for Pauli blocking to find that result for Pauli blocking to find that result.

We have repeated those calculations, both of the cross section from the 
elastic scattering of neutrons from ${}^{12}$C and of the sub-threshold 
spectra for ${}^{13}$C, but now with a view to identify the importance of
each component in the interaction potential. We also study the effects of
the order to which the deformation is taken; specifically we compare results 
found when nuclear surface deformation is taken at $0^{\rm th}$, $1^{\rm st}$, 
and $2^{\rm nd}$ order (the result given in Ref.~\cite{Am03}). 

\begin{figure}
\scalebox{0.6}{\includegraphics*{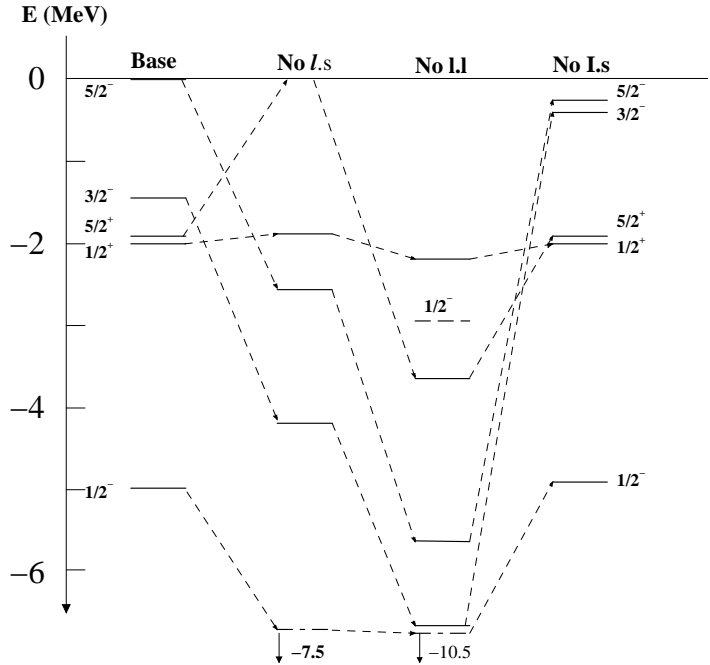}}
\caption{ \label{n-12C-bd}
The spectra of ${}^{13}C$ determined from 
the different calculations described in the text.}
\end{figure}

The effects of the components of the base interaction potential upon the 
sub-threshold (${}^{13}$C) bound states are displayed in Fig.\ref{n-12C-bd}.
In this diagram, and as used throughout, the complete interaction results 
as published previously~\cite{Am03}, are designated as ``Base''.  Those labeled 
by ``No $\ell\cdot s$'' are the results obtained by setting the $V_{\ell s}$ 
parameters to zero leaving the others unaltered.  Likewise those results 
labeled ``No $\ell\cdot \ell$'' and ``No $I\cdot s$'' are ones found
by setting just the $V_{\ell \ell}$ and $V_{Is}$ parameters to zero respectively,
with all others as in Table~\ref{OMparams}. The spectra of ${}^{13}$C determined 
on using the complete interaction is shown at the extreme left.
The three spectra to the right of that are the results obtained when each
of the three operator terms as identified above are omitted from calculation.
%%%%%%%%%%%%%%%%%%%%%%%%%%%%%%%%%%%%%%%%%%%%%%%%%%%%%%%%%%%%%%%%%%%%%%
%%%%%  Figure 1  %%%%%
%%%%%%%%%%%%%%%%%%%%%%%%%%%%%%%%%%%%%%%%%%%%%%%%%%%%%%%%%%%%%%%%%%%%%%
While the $I\cdot s$ component of the interaction is 
a fine tuning effect on the sub-threshold states, the $\ell\cdot s$
and the $\ell\cdot \ell$ components have large influence. 
They have offsetting effects so that the final spectrum has its ground state
at a much lesser (and proper) binding.  The two terms of course influence
different attributes in the spectrum; the spin-orbit governing the splitting of
results for the two values of $j = l \pm \frac{1}{2}$ while the orbit-orbit
controls the splitting of orbits on the basis of their $l$-values.
Relating changes in binding energies to the result 
designated as Base, the excision of the $\ell\cdot s$
component causes an increase in the binding of the negative parity states
by $\sim 2.5$ MeV, reflecting the component of each of those states formed
by coupling of a $p$-wave neutron to ${}^{12}$C. However the $\frac{1}{2}^+$ 
state is but slightly shifted; indicating a weak $d$-wave neutron coupling
in the description of that state.
On the other hand the $\frac{5}{2}^+$ state is markedly shifted into 
the continuum; consistent with it being formed strongly by the coupling of a 
$d$-wave neutron with the $2^+$ state in ${}^{12}$C.
The effect of excising the $\ell\cdot \ell$ term
also is marked.  The negative parity states are  all much more bound; 
this time by $\sim 5.5$ MeV, while again the $\frac{1}{2}^+$ state is but 
slightly changed in binding energy. These effects are consistent with the
dominance of $p$-wave coupling forming the negative parity states while it
is $s$-wave coupling that is predominant with the $\frac{1}{2}^+$ state.
Now however, a second $\frac{1}{2}^-$ state is bound, being brought down from 
the continuum. Finally the $\frac{5}{2}^+$ state again is more bound, once more 
reflecting a strong $d$-wave coupling character.

\begin{figure}
\scalebox{0.75}{\includegraphics*{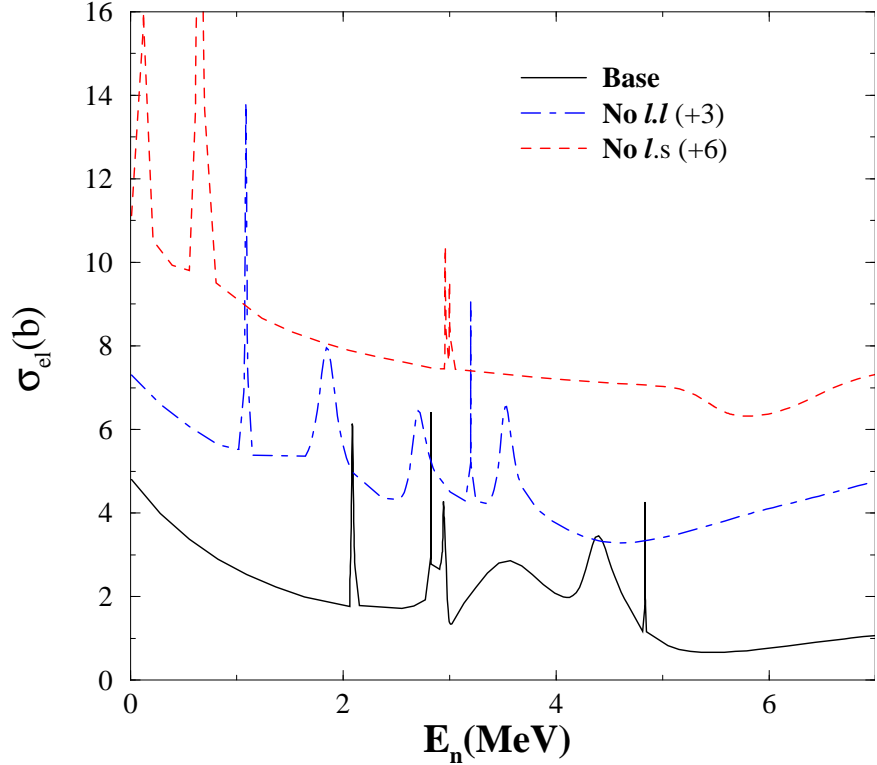}}
\caption{ \label{n-12C-cs}
Total elastic cross sections for n-$^{12}$C scattering as functions
of neutron energy showing the effects of omitting single components of 
the input interaction.}
\end{figure}

The cross sections that result when components of the interaction are excised 
selectively are displayed in Fig.~\ref{n-12C-cs}. 
Therein the restricted potential results have been shifted by 3 and 6~b
as indicated so that the comparisons are clearer.  Omitting the $I\cdot s$
component gives values little different from the Base result (solid curve)
and so they are not shown. 
Note, however, that the limited sensitivity of the cross sections upon the
$I\cdot s$ term was expected since the cross-section values are
dominated by  positive parity amplitudes and the positive parity 
$I\cdot s$
potential strength is small. Also the $I\cdot s$ term does not contribute 
to interactions involving the ground state ($0^+$); interactions that greatly
influence the background scattering.

But the cross sections are very different when the 
$\ell\cdot s$ and the $\ell \cdot \ell$ terms are omitted 
(the dashed and the dot-dashed curves respectively).  
Not only are the resonances shifted in their location but also they have
widths quite different to those found with the base interaction. Again 
the interference between the two operator components is crucial in finding
the result that agrees well with data.

%%%%%%%%%%%%%%%%%%%%%%%%%%%%%%%%%%%%%%%%%%%%%%%%%%%%%%%%%%%%%%%%%
%%%%%  Figure 2  %%%%%
%%%%%%%%%%%%%%%%%%%%%%%%%%%%%%%%%%%%%%%%%%%%%%%%%%%%%%%%%%%%%%%%%%%%%%

\begin{figure}
\scalebox{0.6}{\includegraphics*{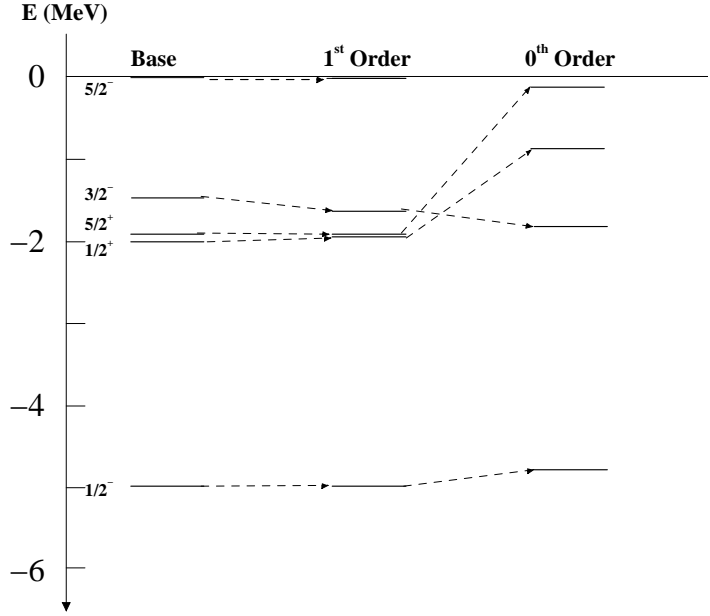}}
\caption{ \label{Spectrum}
The spectra of ${}^{13}C$ determined from 
calculations made on reducing the order of the deformation.}
\end{figure}

The spectra shown in Fig.~\ref{Spectrum} and labeled 
1$^{\rm st}$ order and 0$^{\rm th}$ order, result when deformation 
is taken only to first order or not at all.
The first  we achieved by setting to zero the value of 
$\beta_2^2\ (\propto \kappa^2)$
in the potential matrices of Eq.~B.3 in Appendix B of ref.~\cite{Am03}.
Of course $\beta_2\ (\propto \kappa)$ was taken as -0.52 therein still.
The second resulted from a calculation made using $\beta_2 = 0$ everywhere.
%%%%%%%%%%%%%%%%%%%%%%%%%%%%%%%%%%%%%%%%%%%%%%%%%%%%%%%%%%%%%%%%%%%%%%
%%%%%  Figure 3  %%%%%
%%%%%%%%%%%%%%%%%%%%%%%%%%%%%%%%%%%%%%%%%%%%%%%%%%%%%%%%%%%%%%%%%%%%%%

In 0$^{\rm th}$ order coupling, we obtain  single particle states
of the base potential. Of those, the p-shell terms are 
reasonable when compared to the known spectrum of ${}^{13}$C. 
In ref.~\cite{Am03}, the $p-sd$ model shell structures for these
were listed as dominantly a $p$-shell neutron coupled to the 
ground state of ${}^{13}$C.
Allowing coupling to first order in the deformation compresses the spectrum 
as now the calculations allow for an appreciable, and realistic,
component of $p-sd$-shell particles coupled to the $2^+$ state
of ${}^{12}$C in the description of the states in ${}^{13}$C. 
The spectrum obtained is
very similar to that resulting when deformation is taken to second order
(the Base result). Treating deformation to second order increases
the number of contributing terms quite markedly as is evident when one 
considers the entries in Eq.~(B.3) of ref.~\cite{Am03} that have the scale
quantity $\kappa^2$. But the corrections they make to the sub-threshold 
spectra are relatively minor.  Notably the positive parity pair of states are 
slightly more separated and the $\frac{3}{2}^-$ state slightly less bound.

\begin{figure}
\scalebox{0.75}{\includegraphics*{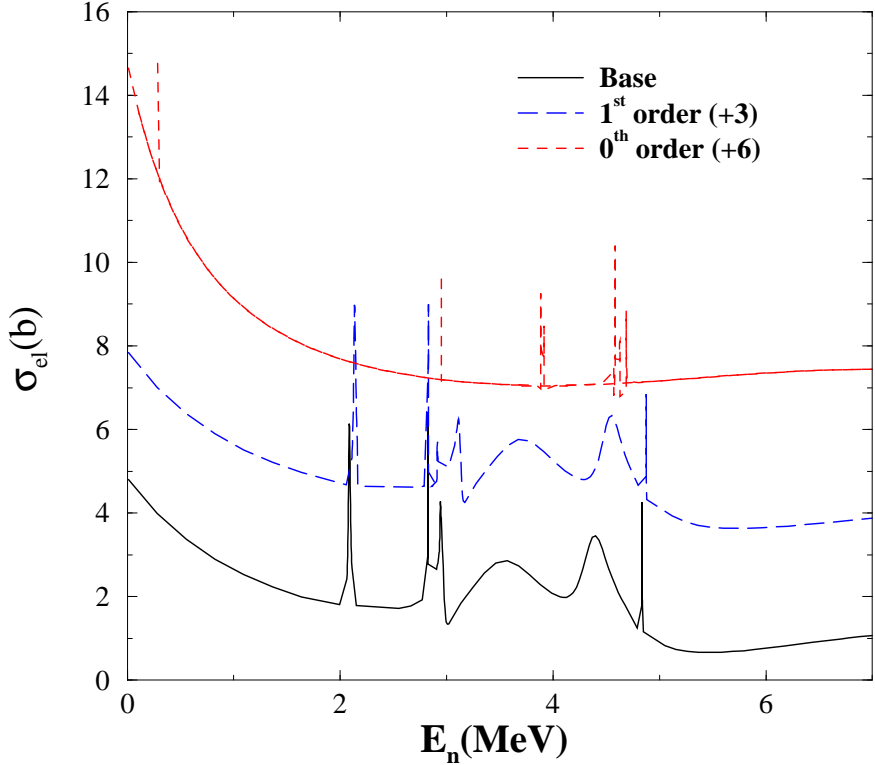}}
\caption{ \label{Order-xsec}
Total elastic cross sections for n-$^{12}$C scattering as functions
of neutron energy showing the effects of reducing the order of
deformation involved.} 
\end{figure}

In Fig.~\ref{Order-xsec} we display the cross sections for 
neutron-${}^{12}$C elastic scattering that result when deformation
is neglected (dashed line), is taken to first order (long-dashed line)
and is taken to second order (the Base result as displayed by the 
continuous line).  Again to add in clarity of viewing, the restricted 
calculation results have been shifted (again by 3 and 6~b) so that the curves 
can more easily be distinguished.
As is evident, there is little to distinguish between the results when
deformation is taken to first and to second order. Overall there is a slight
inward shift of the resonances with energy; the more so with the broader 
$\frac{3}{2}$ resonances in the 3 to 5 MeV region.
But neglect of deformation has a most marked effect. Not only do the narrow
resonances become bound states in the continuum, but also the large shape 
resonances vanish. 
%%%%%%%%%%%%%%%%%%%%%%%%%%%%%%%%%%%%%%%%%%%%%%%%%%%%%%%%%%%%%%%%%%%%
%%%%%  Figure 4  %%%%%
%%%%%%%%%%%%%%%%%%%%%%%%%%%%%%%%%%%%%%%%%%%%%%%%%%%%%%%%%%%%%%%%%%%%%%

\begin{figure}
\scalebox{0.75}{\includegraphics*{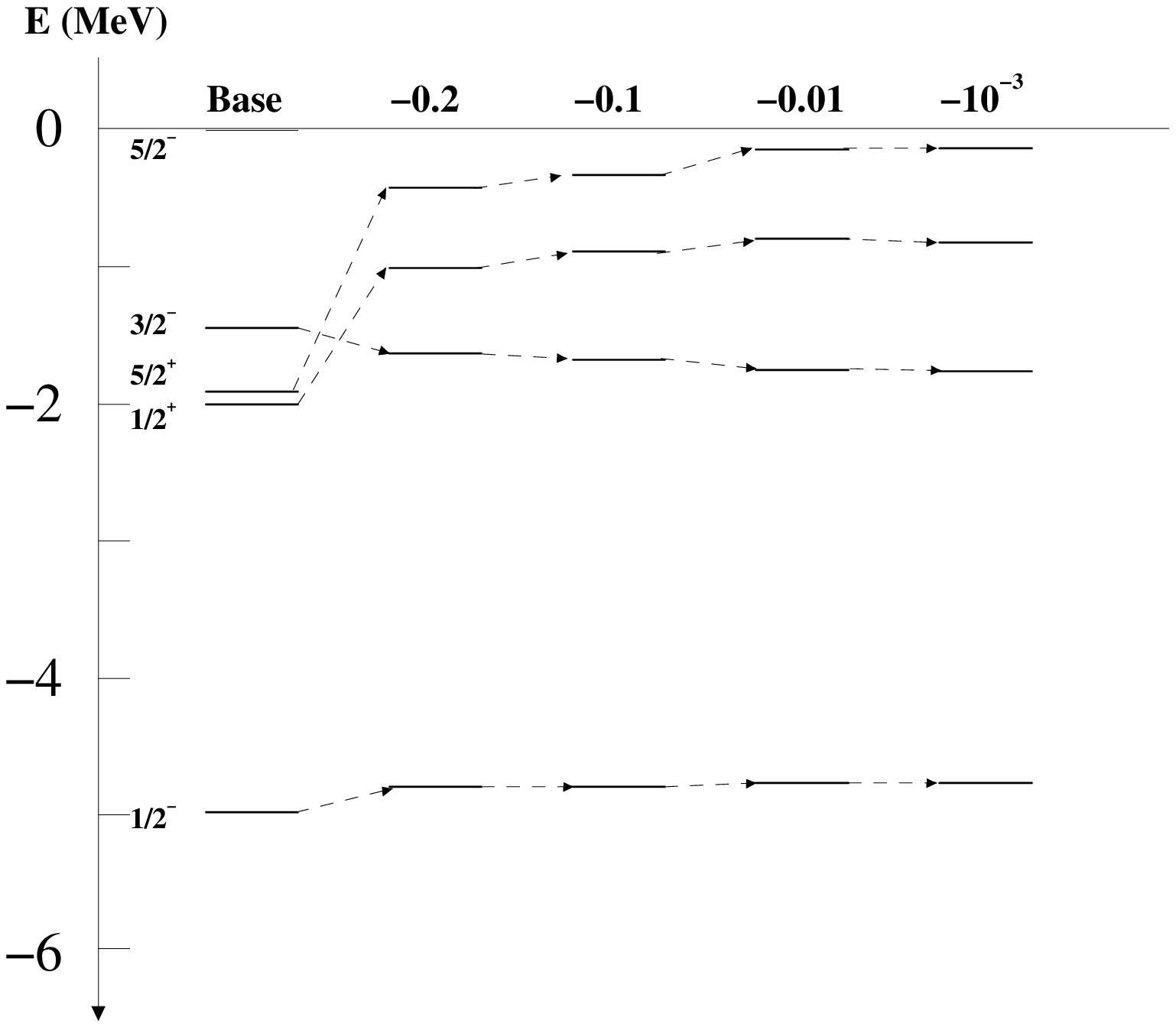}}
\caption{ \label{n-12C-spect-beta}
Spectra of $^{13}$C determined by reducing the value of $\beta_2$
from the Base value (-0.52) to those given above each column.}
\end{figure}

Finally we consider the effects of changes in the value of the deformation
parameter.  Values of -0.52 (Base), -0.2, -0.1, -0.05, -.001, and
of -0.0001 have been used in calculations of both the sub-threshold bound 
states as well as of the scattering cross sections. The bound states 
(of ${}^{13}$C) vary with the values of $\beta_2$ as shown in 
Fig.~\ref{n-12C-spect-beta}. There is a rapid decrease of the energies
of the ${\frac{1}{2}}^+$ and ${\frac{5}{2}}^+$ states as $\beta_2$ is reduced 
to a value of -0.2 and thereafter, those states slowly increase in excitation
above the ${\frac{1}{2}}^-$ ground state.
The two negative parity state, on the other hand, vary little in their
binding with change in $\beta_2$, being slightly compressed from what 
values they have from the base calculation.
%%%%%%%%%%%%%%%%%%%%%%%%%%%%%%%%%%%%%%%%%%%%%%%%%%%%%%%%%%%%%%%%%%%%
%%%%%  Figure 5  %%%%%
%%%%%%%%%%%%%%%%%%%%%%%%%%%%%%%%%%%%%%%%%%%%%%%%%%%%%%%%%%%%%%%%%%%%%%

\begin{figure}
\scalebox{0.75}{\includegraphics*{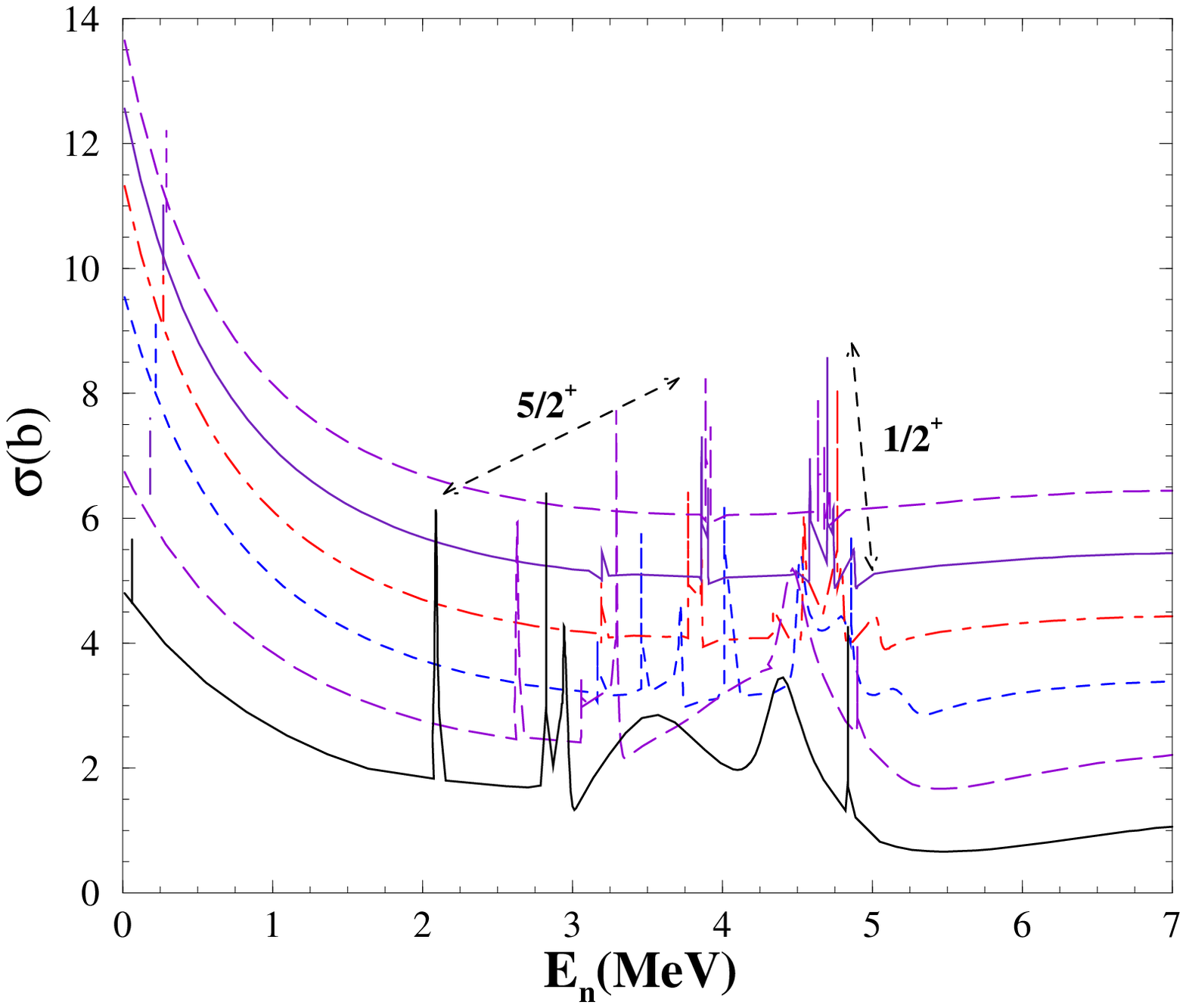}}
\caption{ \label{xsec-beta}
Total elastic cross sections for n-$^{12}$C scattering as functions
of neutron energy showing the effects of reducing the value of
$\beta_2$ as described in the text.}
\end{figure}

The cross sections that result when different $\beta_2$ values 
were used in our calculations are shown in Fig.~\ref{xsec-beta}.
Therein each value is scaled upward overall by a `$n$' barn, where 
$n$ is the number of the sequence above the Base result (solid curve) 
as follows: the lower long-dashed curve is the result on using $\beta_2 = -0.4$,
that displayed by the small dashed curve had $\beta_2 = -0.2$,
the dot-dashed curve portrays the result with $\beta_2 = -0.1$ 
the next (a solid curve) was found on using $\beta_2 = -0.05$, while the topmost
(long dashed) curve is the cross section found with $\beta_2 = -0.01$.
The general trend is that the background cross section values near 
threshold increase in size and that is consistent with the strong
sub-threshold $s$-wave ${\frac{1}{2}}^+$ state moving closer to threshold
as $\beta_2$ decreases. Then what was just a sub-threshold state, the 
${\frac{5}{2}}^-$ in the Base calculation moves into the positive energy regime
with decrease in $\beta_2$. That state has an extremely small width
($\le 10^{-11} MeV$) so that its existence is known but the strength
has not been ascertained.  An arbitrary cross section spike has been added
to these results at the known energies ($\le$ 0.3 MeV).
Then as $\beta_2$ decreases, while some resonance features of the cross section 
move to higher excitation, notably the $\frac{5}{2}_1^+, 
\frac{3}{2}_1^+, \frac{3}{2}_2^+, {\rm and}\ \frac{7}{2}_1^+$
resonances, others essentially remain unchanged in energy.
That is indicated in the figure by the dashed lines showing the 
energy movement of the $\frac{5}{2}_1^+$ and $\frac{1}{2}_1^+$ resonances.
Also the broad $\frac{3}{2}^+$ (shape) resonances
shrink until, with $\beta = 0$ they disappear.
%%%%%%%%%%%%%%%%%%%%%%%%%%%%%%%%%%%%%%%%%%%%%%%%%%%%%%%%%%%%%%%%%%%%
%%%%%  Figure 6  %%%%%
%%%%%%%%%%%%%%%%%%%%%%%%%%%%%%%%%%%%%%%%%%%%%%%%%%%%%%%%%%%%%%%%%%%%%%

Summarizing, the MCAS approach to analyze (low-energy) nucleon-nucleus
scattering is built from a model structure of the interaction potentials
between a nucleon and each of the target states taken into consideration.
The process finds all resonances (narrow and broad) in the selected 
(positive) energy range, as well as defining the background cross section.
The method also can be used with negative energies to specify the
sub-threshold bound states of the compound nucleus.

With an interaction matrix of potentials defined by a collective
model of the structure of the ground ($0^+$), $2^+$ (4.44 MeV), and $0^+_2$ 
(7.65 MeV) states in ${}^{12}$C, and with deformation therein taken to 
second order, MCAS calculations of neutron scattering have been made with 
the complete (Base) results well matching the observed scattering 
data and producing sub-threshold bound states very like those known 
in the spectrum of ${}^{13}$C.  Note however, as stressed before~\cite{Am03},
allowance for the influence of the Pauli principle was  crucial to 
that achievement.  
Once the non-locality due to Pauli blocking was adequately treated,
good agreement with respect to sub-threshold spectra and scattering
cross sections were obtained with a phenomenological (coupled-channel)
interaction. However, in addition to usual central and
$s\cdot \ell$ (spin-orbit) potentials, we have found that additional operator
terms of the type $\ell \cdot \ell$ (orbit-orbit) and $s\cdot I$ 
(spin-spin) were necessary. These latter terms provide contributions to
the usual interaction, particularly for negative parity, to get the appropriate
separation energies of sub-threshold bound and resonance states.    

In addition we have shown how the results
depend on the coupling deformation parameter and,
further, we have shown that both the spectrum (of ${}^{13}$C)
as well as the specific resonance structure of the cross sections converge
well when the collective model interactions are taken 
to second order in that deformation. 

%%%%%%%%%%%%%%%%%%%%%%%%%%%%%%%%%%%%%%%%%%%%%%%%%%%%%%%%

\bibliography{n-12C-comp}

\end{document}